\documentclass[epj,referee]{svjour}
\usepackage{latexsym}
\usepackage{graphics}
\usepackage{amsmath}
\usepackage{amssymb}
\begin{document}
\title{High frequency magnetic oscillations of the organic metal $\theta$-(ET)$_4$ZnBr$_4$(C$_6$H$_4$Cl$_2$) in pulsed magnetic field of up to 81 T}
\titlerunning{High frequency magnetic oscillations up to 81 T}
\author{J. B\'{e}ard\inst{1} \and J. Billette\inst{1} \and M. Suleiman\inst{1} \and P.~Frings\inst{1}
\and W. Knafo\inst{1} \and G. W. Scheerer\inst{1} \and F.~Duc\inst{1} \and D.~Vignolles\inst{1} \and
M. Nardone\inst{1} \and A.~Zitouni\inst{1} \and P.~Delescluse \inst{1} \and
J.-M. Lagarrigue\inst{1} \and F.~Giquel\inst{1} \and B. Griffe\inst{1} \and
N. Bruyant\inst{1} \and J.-P. Nicolin\inst{1} \and G.~L.~J.~A.~Rikken\inst{1} \and R.~B.~Lyubovskii\inst{2} \and G. V. Shilov\inst{2} \and E. I. Zhilyaeva\inst{2}
\and R.~N.~Lyubovskaya\inst{2} \and A. Audouard\inst{1}\thanks{\emph{Electronic address:} alain.audouard@lncmi.cnrs.fr} 
\authorrunning{J. B\'{e}ard et al}
}                     
%
%
\institute{
\inst{1} Laboratoire National des Champs Magn\'{e}tiques
Intenses (UPR 3228 CNRS, INSA, UJF, UPS) 143 avenue de Rangueil,
F-31400 Toulouse, France.\\
\inst{2} Institute of Problems of Chemical Physics, RAS, 142432 Chernogolovka, MD, Russia.\\
}
\date{Received: \today / Revised version: date}
%
\abstract{
De Haas-van Alphen oscillations of the organic metal $\theta$-(ET)$_4$ZnBr$_4$(C$_6$H$_4$Cl$_2$) are
studied in pulsed magnetic fields up to 81 T. The long decay time of the pulse allows determining
reliable field-dependent amplitudes of Fourier components with frequencies up to several kiloteslas. The Fourier
spectrum is in agreement with the model of a linear chain of coupled orbits. In this model, all the observed frequencies are linear
combinations of the frequency linked to the basic orbit $\alpha$ and to the magnetic-breakdown orbit $\beta$. } 
\maketitle
\section{Introduction}
\label{intro}
High magnetic fields produced by pulsed magnets are known to be powerful tools for solid state physics.
Among the experiments performed at very high magnetic fields (above 70~T), right from the late eighties, single-turn destructive coils have been successfully used for the study of metamagnetic transitions in the quasi-one dimensional Ising antiferromagnet CsCoCl$_3$ up to 85 T \cite{Ta88} and, shortly afterwards, up to 94 T in the itinerant paramagnet YCo$_2$ \cite{Go89}. Nevertheless, such pulses have very short durations of a few $\mu$s, hampering reliable determination of phenomena involving abrupt field-dependent changes of any physical property.

\begin{figure}
\centering
\resizebox{0.9\columnwidth}{!}{%
  \includegraphics{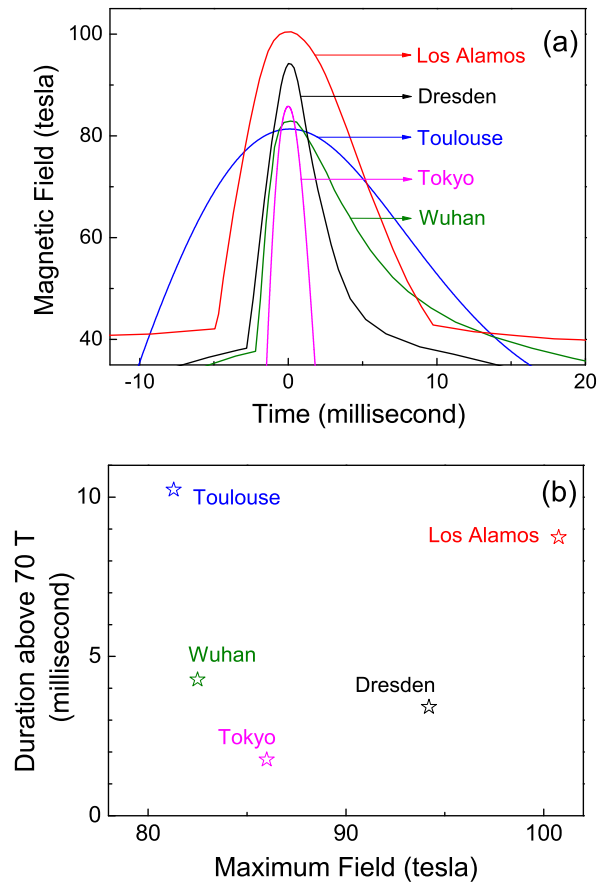}
}
\caption{(Color on line) (a) Temporal profiles of the highest non-destructive magnetic field pulses generated at different pulsed-field facilities
(b) Duration time above 70~T versus maximum magnetic field of these pulses.}
\label{fig:comp_profiles}       
\end{figure}

In recent years, efforts have been made to develop non-destructive devices providing fields above 70~T with longer pulses. The most recent world record has been established in March 2012 at the National High Magnetic Field Laboratory (NHMFL)-Los Alamos \cite{Se12}, with a non-destructive pulse of up to 100.75~T. Competition between the different pulsed field facilities is intense, and record non-destructive fields of 94.2~T, 86~T, and 82.5~T, were recently obtained at the HLD-Dresden \cite{Dresden}, the ISSP-Tokyo \cite{Tokyo}, and the WHMFC-Wuhan \cite{WuHan}, respectively. However, one should always bear in mind that these magnets are research tools, and their utility depends not exclusively on the field strength. Other factors, like pulse duration and bore size are essential for the realization of state-of-the-art high field measurements. Part of the progress obtained with the recent record fields was at the expense of these other parameters, limiting the usefulness of such magnets for experiments. Within this context, the LNCMI-Toulouse pulsed field facility strategy is to focus on offering optimal experimental conditions closest to record field strengths. The newest double coil system recently put into operation in Toulouse 'only' generates magnetic fields up to 81.2 T, but it offers a world record field duration of 10.2 ms above 70~T, reducing eddy current problems and facilitating high quality data acquisition (see Fig.~\ref{fig:comp_profiles}).

Up to now, experimental studies in non-destructive magnetic fields higher than 80~T have only been reported at the NHFML-Los Alamos. Recent examples are the determination of a cascade of metamagnetic transitions in LaCoO$_3$ \cite{Al12} and the observation of quantum oscillations with frequencies of few hundredth of teslas \cite{Se12} in pulsed fields up to 97 T and 100.75 T, respectively. We present here the first experimental study performed in Europe in magnetic fields higher than ~80~T. De Haas-van Alphen (dHvA) oscillations of the quasi-two-dimensional organic metal $\theta$-(ET)$_4$ZnBr$_4$(C$_6$H$_4$Cl$_2$), where ET stands for the bis(ethylenedithio)tetrathiafulvalene molecule, are studied up to 81 T at 1.5 K. Indeed, as reported hereafter, its oscillatory spectrum involves frequencies as high as several kiloteslas. In line with statements of ~\cite{Au12}, field-dependent amplitudes of the studied basic and magnetic-breakdown (MB) orbits are in agreement with the Lifshits-Kosevich and Falicov-Stachowiak models \cite{Sh84} in a large field range. We demonstrate that reliable determination of physical properties rapidly varying with magnetic field can be obtained with the new long-pulse 80-T magnet of the LNCMI-Toulouse, since the decay time of the pulse is long enough.

\section{Experimental}

\begin{figure}
\centering
\resizebox{0.9\columnwidth}{!}{%
  \includegraphics{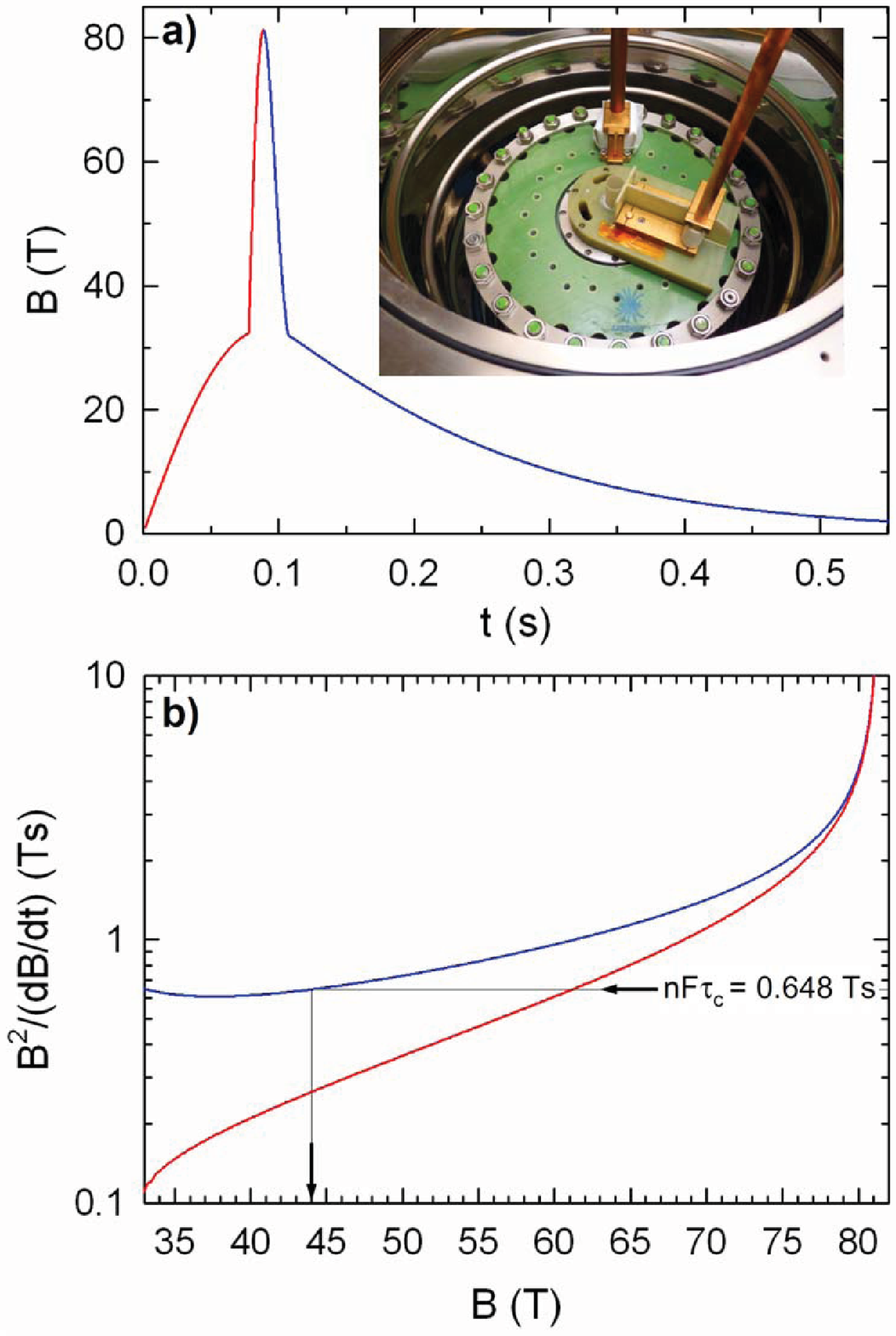}
}
\caption{(Color on line) (a) Temporal profile of the pulsed field. The inset in (a) displays the top of the two-coil device. (b) Field dependence of the parameter B$^2$/(dB/dt) relevant to the high field part. Horizontal line marks the $nF\tau_c$ value corresponding to $n$=8, $F$= 8.1 kT and $\tau_c$ = 10 $\mu$s (see text). The resulting minimum field (44 T) at which reliable data is obtained for this frequency is marked by the vertical arrow. Red and blue solid lines stand for the
raising and decaying part of the pulse, respectively.}
\label{fig:Tc}       
\end{figure}

The studied crystal was synthesized by the electrocrystallization technique reported in \cite{Sh11}. Magnetic torque was measured with a piezoresistive microcantilever as reported in~\cite{Au12}. X-ray diffraction data were collected at 100 K with a KM-4 single-crystal diffractometer (Kuma Diffraction) at the IPCP of Chernogolovka and at 180 K with an Xcalibur diffractometer (Oxford Diffraction) at the Laboratoire de Chimie de Coordination of Toulouse. Pulsed magnetic fields results were obtained with a two-coil system \cite{Fr08,Pe10} in which the external coil provides a background field of 32.3 T. At a time close to the maximum of the background field, the inner coil is powered producing a maximum field of 81.2 T (see Fig.~\ref{fig:Tc}). Both coils are powered by an independent generator. The decay time elapsed between 81.2 T and 32.3 T is 18 ms, i.e. about a factor of two larger than in Ref.~\cite{Se12}.

The time constant($\tau_c$) of the lock-in used for measurements must be large enough to achieve a suitable signal-to-noise ratio. However, reliable data are only obtained provided $\tau_c$ is small enough compared to the temporal variation of the measured quantity (see $e.g.$ \cite{SR830}).  For dHvA oscillations, that are periodic in 1/$B$ with a given frequency F, the time ($\delta t$) elapsed during one oscillation is, at first order, given by $\delta t$ = $B^2/(F|dB/dt|)$. We have empirically checked that the condition $\tau_c < \delta t/n$ with n $\simeq$ 8 must be fulfilled to get reliable value of the oscillation amplitude. Namely experiments performed with time constant such as $\tau_c > \delta t/$8 yield damped oscillations.  The field dependence of $B^2/|dB/dt|$, which should be higher than $nF\tau_c$, is displayed in Fig.~\ref{fig:Tc}b.  As reported in Ref.~\cite{Dr10}, even though $|dB/dt|$ roughly exponentially decreases as the magnetic field decreases, $\delta t$ still decreases as the magnetic field decreases which hamper reliable determination of the oscillations amplitude at low field. As an example, the highest frequency considered in our data analysis (see next section) is $F$ = 8.1 kT. For  $n$ = 8 and $\tau_c$ = 10 $\mu$s, which is the time constant value used for the reported measurements, reliable data relevant to this frequency are obtained for $B^2/(F|dB/dt|)$ $>$  0.648 Ts, $i.e.$, for a magnetic field higher than 44 T (see Fig.~\ref{fig:Tc}b). We emphasize that for (dB/dt) twice as large as for the pulse of Fig.~\ref{fig:Tc}, reliable data regarding this Fourier component would be obtained in a much smaller field range, ranging from 68 T to 81 T, only. Obviously, reliable data relevant to lower frequencies can be obtained down to smaller fields.

\section{Results and discussion}

\begin{figure}
\centering
\resizebox{0.9\columnwidth}{!}{%
  \includegraphics{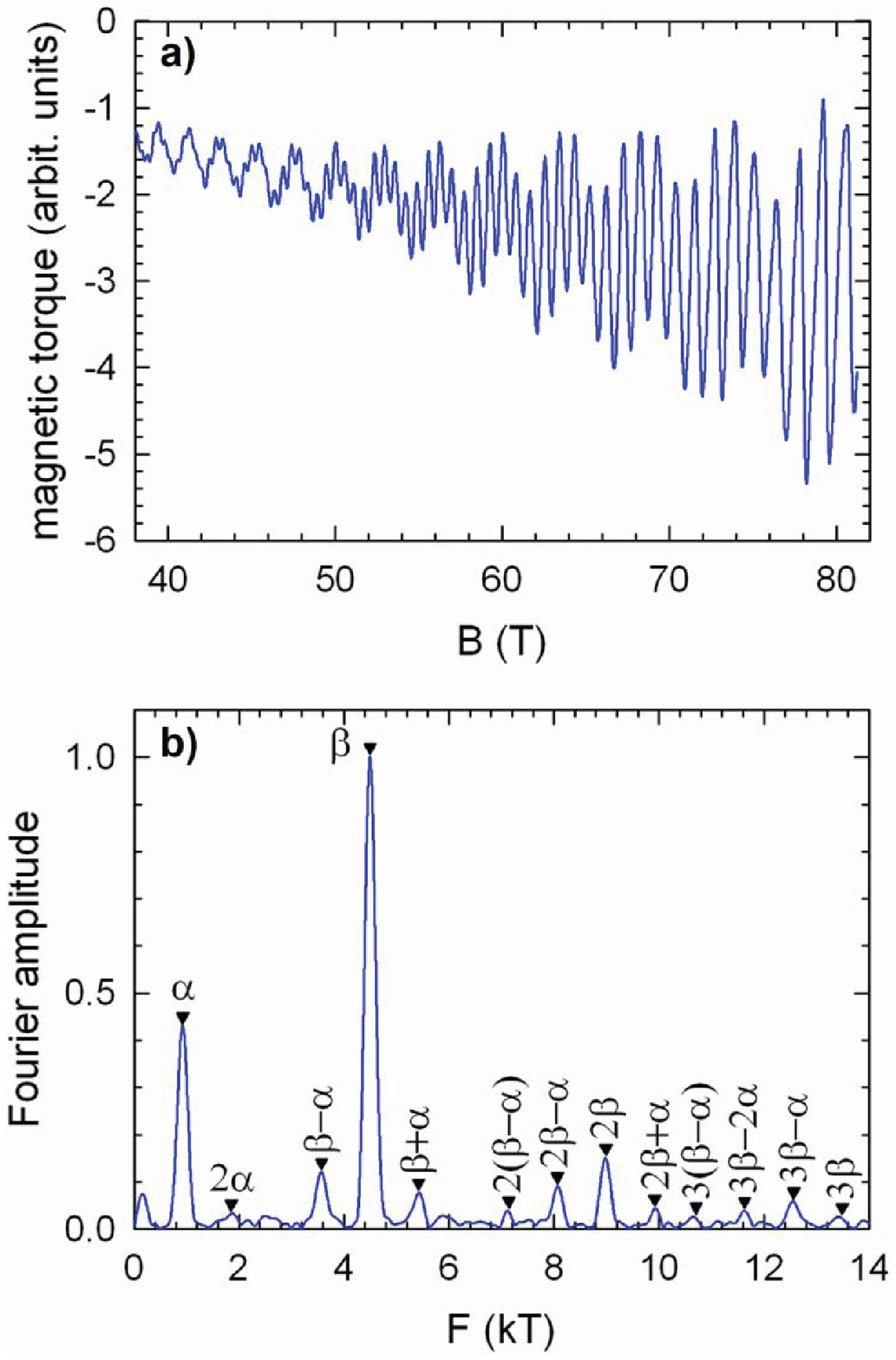}
}
\caption{(Color on line) (a) Magnetic torque at 1.5 K in the high field range and (b) corresponding Fourier analysis.
Solid triangles are marks calculated with F$_{\alpha}$ = 0.93 kT and F$_{\beta}$ = 4.5 kT.}
\label{fig:TF}       
\end{figure}

\begin{figure}
\centering
\resizebox{0.9\columnwidth}{!}{%
  \includegraphics{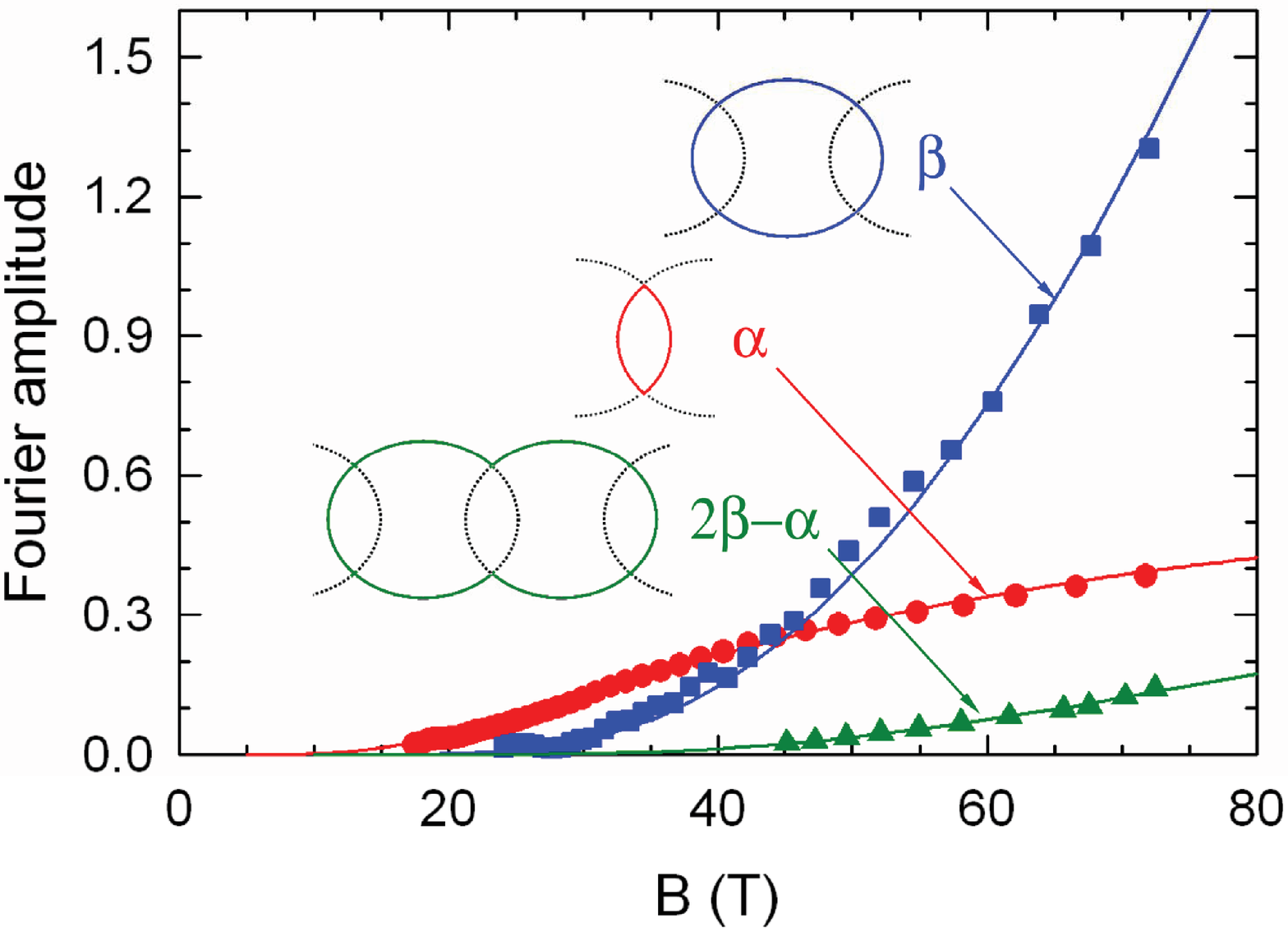}
}
\caption{(Color on line) Field-dependent amplitude of the Fourier components relevant to $\alpha$ (circles),
$\beta$ (squares) and $2\beta-\alpha$ (triangles). Solid lines are obtained from the Lifshits-Kosevich
model with $m_\alpha$ = 1.9, $m_\beta$ = 3.5, $T_D$ = 1.2 K and $B_0$ = 25 T (see text).}
\label{fig:A}       
\end{figure}

Magnetic torque data and the corresponding Fourier analysis in the field range above 44 T are displayed in Fig.~\ref{fig:TF}. All the observed frequencies are linear combinations of F$_{\alpha}$ = 0.93 kT and  F$_{\beta}$ = 4.5 kT. Such a spectrum, relevant to the linear chain of coupled orbits model \cite{Sh84,Pi62}, is in agreement with the data reported for the charge transfer salt $\theta$-(ET)$_4$CoBr$_4$(C$_6$H$_4$Cl$_2$), which differs from the studied compound by substitution of Zn by Co. This result is not surprising since, according to X-ray diffraction data \cite{Au12}, these two compounds are isostructural. For this kind of Fermi surface, F$_{\alpha}$ and F$_{\beta}$ correspond to the closed orbit $\alpha$ and the MB orbit $\beta$, respectively (see Fig.~\ref{fig:A}). In agreement with X-ray diffraction data, the area of the  $\beta$ orbit is equal to that of the first Brillouin zone area.

As discussed in Ref. \cite{Au12}, oscillation spectra can be strongly influenced by both the MB-induced formation of Landau bands and oscillation of the chemical potential, which are liable to induce Fourier components corresponding to 'forbidden orbits'. This is for example the case of $\beta-\alpha$ which cannot correspond to a MB orbit. Besides, field and temperature dependence of several other Fourier components (such as 2$\alpha$ and $\beta+\alpha$) are not in agreement with the semiclassical model of Falicov-Stachowiak \cite{Sh84}. Analysis of these phenomena, which requires a comprehensive field and temperature dependence study, is beyond the scope of this paper. In line with the discussion of the preceding section, only the amplitude of Fourier components with frequencies of at most 8.1 kT can be reliably measured in the field range down to 44 T. In other words, the amplitude of all the Fourier components beyond 2$\beta-\alpha$ are underestimated in the data of Fig.~\ref{fig:TF}b. Therefore, only the basic $\alpha$ and MB-induced $\beta$ and 2$\beta-\alpha$ orbits, which are expected to follow the semiclassical model \cite{Au12}, are considered in the following.

For magnetic torque oscillations, the amplitude $A(\eta)$ of the Fourier component relevant to a given orbit $\eta$ can be written as:

\begin{equation}
\label{eq:LK}
A(\eta) \propto B R_T(\eta) R_D(\eta) R_{MB}(\eta)
\end{equation}

According to the Lifshits-Kosevich and Falicov-Stachowiak models \cite{Sh84}, thermal, Dingle and MB damping factors are given by $R_T(\eta)$ = $u_0Tm_{\eta}/[B sinh(u_0Tm_{\eta}/B)]$, $R_D(\eta)$ = $exp(-u_0T_Dm_{\eta}/B)$ and $R_{MB}(\eta)$ = $p^{n_t(\eta)}q^{n_b(\eta)}$, respectively, where $u_0$ = 2$\pi^2m_ek_B/e\hbar$ and $m_{\eta}$ is the effective mass. In the framework of the Falicov-Stachowiak model, the effective mass of 2$\beta-\alpha$ is given by $m_{2\beta-\alpha}$ = 2$m_{\beta}$-$m_{\alpha}$. The Dingle temperature, $T_{D}$ = $\hbar$/2$\pi$$k_B\tau$ where $\tau$ is the
relaxation time, is assumed to be the same for all the considered orbits. Integers $n^t(\eta)$ and $n^b(\eta)$ are the number of MB tunnelings and Bragg reflections, respectively. The MB tunneling and reflection probabilities are given by $p$ = $\exp(-B_0/2B)$ and $q^2$ = 1 - $p^2$, respectively, where $B_0$ is the MB field. The field-dependent amplitude of the considered orbits is displayed in Fig.~\ref{fig:A}. Solid lines in this figure are calculated with Eq.~\ref{eq:LK}, assuming $m_\alpha$ = 1.9, $m_\beta$ = 3.5, $T_D$ = 1.2 K and $B_0$ = 25 T. Obviously, accurate determination of these parameters requires comprehensive analysis of the temperature dependence of the relevant amplitudes. Nevertheless, a good agreement between experimental data and Eq.~\ref{eq:LK} is observed.

\section{Conclusion}

De Haas-van Alphen oscillations of the organic metal $\theta$-(ET)$_4$ZnBr$_4$(C$_6$H$_4$Cl$_2$) have been measured at 1.5 K in pulsed magnetic field of up to 81 T. The Fourier
spectrum is in agreement with the model of a linear chain of coupled orbits for which all the observed frequencies are linear
combinations of the frequencies linked to the basic orbit $\alpha$ and to the magnetic-breakdown orbit $\beta$. Thanks to the long decay time of the pulse, reliable data relevant to the field-dependent amplitude of the Fourier components with frequencies up to 8.1 kT are obtained for fields above 44 T. Besides, the present article is the first publication of experimental data in non-destructive magnetic fields higher than 80 T obtained in Europe.

\begin{acknowledgement}
 This work has been supported by EuroMagNET II under the EU contract number 228043 and by the CNRS-RFBR cooperation under the PICS contract number 5708. We acknowledge the technical help of Laure Vendier at the X-ray facility of the LCC-Toulouse.
\end{acknowledgement}


\begin{thebibliography}{}
%
%

\bibitem{Ta88} S. Takeyama, K. Amaya, T. Nakagawa, M. Ishizuka, K. Nakao, T. Sakakihara, T. Goto, N. Miura, Y. Ajiro, and H. Kikuchi, J. Phys. E.: Sci. Instrum. \textbf{21} (1988) 1025.

\bibitem{Go89} T. Goto, K. Fukamichi, T. Sakakibara and H. Komatsu, Sol. State Commun. \textbf{72} (1989) 945.

\bibitem{Se12} S.E. Sebastian, N. Harrison, R. Liang, D.A. Bonn, W.N. Hardy, C.H. Mielke and G.G. Lonzarich, Phys. Rev. Lett. \textbf{108} (2012) 196403.

\bibitem{Dresden} Unpublished, see web site: $www.hzdr.de$

\bibitem{Tokyo} K. Kindo, J. Phys.:Conf. Ser. \textbf{51} (2006) 522.

\bibitem{WuHan} Unpublished, see web site: $http://eng.whmfc.cn/$

\bibitem{Al12} M.M. Altarawneh, G.-W. Chern, N. Harrison, C.D. Batista, A. Uchida, M. Jaime, D.G. Rickel, S.A. Crooker, C.H. Mielke, J.B. Betts, J.F. Mitchell, and M.J.R. Hoch, Phys. Rev. Lett. \textbf{109} (2012) 037201.



\bibitem{Au12} A. Audouard, J.-Y. Fortin, D. Vignolles, R. B. Lyubovskii, L. Drigo,
	F. Duc, G. V. Shilov, G. Ballon, E. I. Zhilyaeva, R. N. Lyubovskaya
	and E. Canadell, EPL \textbf{97} (2012) 57003.

\bibitem{Sh84} D. Shoenberg, \textit{Magnetic Oscillations in Metals} (Cambridge University Press, Cambridge, 1984).

\bibitem{Sh11} G. V. Shilov, E. I. Zhilyaeva, A. M. Flakina, S. A. Torunova, R. B. Lyubovskii, S. M. Aldoshin, and R. N. Lyubovskaya, Cryst. Eng. Comm.\textbf{13} (2011) 1467.

\bibitem{Fr08} P. Frings, J. Billette, J. B\'eard, O. Portugall, F. Lecouturier and G. Rikken, IEEE Trans. Appl. Supercond. \textbf{18} (2008) 592.

\bibitem{Pe10} J.A.A.J. Perenboom, P. Frings, J. B\'eard, B. Bansal, F. Herlach, Tao Peng and S. Zherlitsyn,  J. Low Temp. Phys. \textbf{159} (2010) 336.

\bibitem{SR830} Model SR830 DSP lock-in amplifier instruction manual (Sunnyvale, Standford Research Systems Inc., 2011)

\bibitem{Dr10} L. Drigo, F. Durantel, A. Audouard and G. Ballon, Eur. Phys. J.-Appl. Phys. \textbf{52} (2010) 10401.

\bibitem{Pi62} A. B. Pippard, Proc. Roy. Soc. (London) \textbf{A270} (1962) 1.






\end{thebibliography}
%

\end{document}